# The Geant4-DNA project

S. Incerti[1,2,3], G. Baldacchino[1,4], M. Bernal[1,5], R. Capra[1], C. Champion[1,6], Z. Francis[1,7], S. Guatelli[1,8], P. Guèye[1,9], A. Mantero[1,2,3], B. Mascialino[1,10], P. Moretto[1,2,3], P. Nieminen[1,11], A. Rosenfeld[1,8], C. Villagrasa[1,7] and C. Zacharatou[1,12]

[1] The Geant4-DNA collaboration
[2] CNRS/IN2P3, Centre d'Etudes Nucléaires de Bordeaux-Gradignan, UMR 5797, Gradignan F-33175, France
[3] Université de Bordeaux, Centre d'Etudes Nucléaires de Bordeaux-Gradignan, UMR 5797, Gradignan F-33175, France
[4] CEA, DSM, IRAMIS, SCM, Laboratoire de Radiolyse, Saclay, France
[5] Department of Physics, Simon Bolivar University, P.O. Box 89000, Caracas 1080-A, Venezuela
[6] Université Paul Verlaine-Metz, Laboratoire de Physique Moléculaire et des Collisions, 1 Boulevard Arago, Technopôle 2000, 57078 Metz, France
[7] IRSN, Institut de Radioprotection et de sureté Nucléaire, BP17, 92262 Fontenay-aux-Roses France
[8] Centre for Medical Radiation Physics (CMRP), University of Wollongong, NSW, Australia
[9] Hampton University, Physics Department, Hampton VA 23668, USA
[10] Karolinska Institutet, Box 260, S-171-76, Stockholm, Sweden
[11] ESA-ESTEC, 2200 AG Noordwjik, The Netherlands
[12] Department of Radiation Oncology, Copenhagen University Hospital, Rigshospitalet, Blegdamsvej 9, DK-2100 Copenhagen, Denmark

**Abstract**—The Geant4-DNA project proposes to develop an open-source simulation software based and fully included in the general-purpose Geant4 Monte Carlo simulation toolkit. The main objective of this software is to simulate biological damages induced by ionising radiation at the cellular and sub-cellular scale. This project was originally initiated by the European Space Agency for the prediction of deleterious effects of radiation that may affect astronauts during future long duration space exploration missions. In this paper, the Geant4-DNA collaboration presents an overview of the whole on-going project, including its most recent developments already available in the last Geant4 public release (9.3 BETA), as well as an illustration example simulating the direct irradiation of a chromatin fibre. Expected extensions involving several research domains, such as particle physics, chemistry and cellular and molecular biology, within a fully interdiciplinary activity of the Geant4 collaboration are also discussed.

## 1 Introduction

### 1.1 Context

Being able to understand and simulate adverse effects of ionising radiation at the cellular and sub-cellular scale remains a challenge of today's radiobiology research. In particular, validated simulation tools are nowadays of primary importance for human radioprotection in specific professional activities.

Radioprotection is a real concern on a routine basis for example to nuclear power plant workers, to health-care personnel, to particle accelerator personnel working for industrial companies and research laboratories, and even to the general public through indoor exposure to radon from soil emission. They are all continuously exposed to low doses of radiation (the "low dose" regime is of the order of a few μSv). In this regime, the estimation of health hazards from exposure to ionising radiation is limited because no experimental data are available. For the moment, biological effects resulting from such low-dose radiation exposure can only be extrapolated from data that were collected at much higher doses of radiation, for example from the epidemiological surveys performed after the Hiroshima and Nagasaki bombings [1]. It is agreed as a general consensus that biological effects resulting from radiation are proportional to the absorbed dose and consequently that potentially radiation exposure at any dose level can lead to a biological effect, with no dose threshold. Consequently, the risks to human health from low dose radiation exposure remain largely unresolved.

In the case of radiations characterised by high RBE, the availability of simulation tools able to model biological effects of radiation is becoming pivotal, due to the spread of new therapeutic techniques based on ionising radiation, such as ion therapy. Adopting these new techniques, malignant cells are irradiated with heavy particles, such as carbon ions, in order to improve target coverage as well as the biological effectiveness of the treatment, in comparison to traditional treatments. A similar need for simulation tools in the high equivalent dose regime is becoming more and more obvious, as national space agencies are putting huge efforts in the preparation of long duration manned space exploration missions. In particular, long journeys aboard the International Space Station occur more frequently, and the future exploration missions with astronauts towards the planet Mars foresee exposure to space radiation for more than a year, considering the crossing of the Van Allen radiation belts, the transit far from the Earth's magnetosphere and during the stay on the planet. Consequently, simulation tools able to predict not only the dose profiles deposited in cells by the primary and secondary particles, but also the resulting biological damages are urgently needed.

A large experimental and modelling activity is currently taking place, aimed at better understanding the biological effects of ionising radiation at the sub-cellular scale. A considerable amount of experimental data have been accumulated over the past decades in order to measure quantities such as macroscopic cellular survival curves and DNA strand damages after irradiation, in particular with the development of microbeam facilities, which allow a precise targeting of individual cells and control of the delivered dose. Numerous semi-empirical and often complex models based on these experimental data have been proposed, with the aim of building a deterministic theory of cell survival. Such models, e.g. the linear-quadratic model [2], are able to predict the survival of cells as a function of the absorbed dose (among the other variables). In parallel, computer codes have proposed to use a stochastic approach to model physical interactions and the corresponding energy deposits occurring in the irradiated medium, in order to predict subsequent sub-cellular damages (see for example a list of Monte Carlo codes in [3]). These "track structure codes", have been developed for microdosimetry simulations. They are able to simulate precisely particle-matter interactions, the so-called "physical stage of the cellular response to radiation, some of them including the "physico-chemical" and "chemical" stages taking place after the physics stage, and allowing in particular the simulation of the production and tracking of oxidative radical species, including their mutual interactions. With the use of sophisticated geometry models, some of these programs are even able to predict with a reasonable precision direct and non-direct biological damages to the DNA molecule, like the PARTRAC software [4], one of the most complex software developed so far. However, these codes have in general limited application scopes as they can be used for very specific applications, while others are not made publicly available. Moreover, it is in general difficult to use them within the most recent computing environments and tools available today.

The presented project – named "Geant4-DNA" – proposes to develop an experimentally validated simulation platform for the modelling of DNA damage induced by ionising radiation, with the help of modern computing tools and techniques. This is an ambitious work of highly interdisciplinary nature, involving several research fields and gathering experts in elementary particle physics, chemistry, biophysics, molecular and cellular biology, and computer scientists. The platform will be based on the general-purpose and open-source "Geant4" Monte Carlo simulation toolkit, and will benefit from of the toolkit's full transparency and free availability [5]-[7].

### 1.2 The general-purpose Monte Carlo and open-source approaches

The Geant4-DNA project proposes to adopt the Monte Carlo approach. Although Monte Carlo simulation methods may require large amounts of computing power in order to simulate complex situations and setups, they may reach high accuracy and might be considered as an alternative solution to deterministic approaches. Monte Carlo methods make use of random number generators in order to reproduce the stochastic nature of physical interactions occurring between particles and matter.

Instead of developing a dedicated Monte Carlo software with a scope limited to microdosimetry and the simulation of biological damage of radiation only, the Geant4-DNA project aims to extend the general-purpose Geant4 Monte Carlo simulation toolkit with such capabilities. The Geant4 toolkit is a state-of-the-art simulation toolkit describing particle-matter interactions. It was originally developed at the European Organization for Nuclear Research (CERN) in Geneva, Switzerland, for the simulation of High Energy Physics (HEP) experiments at the Large Hadron Collider (LHC). The development of Geant4 adopts the open-source strategy: the software is entirely transparent and totally free. Anyone can download the Geant4 toolkit and develop his/her own simulation applications. The toolkit is being developed by a team of more than 80 international collaborators and worldwide users are actively involved in its validation, aimed at determining its accuracy at both microscopic and macroscopic levels. Public releases are available twice a year.

The Geant4 software uses the object-oriented technology (C++) providing remarkable flexibility and extensibility, which progressively lead to the development of novel Geant4 applications in research domains involving particle-matter interactions not limited to High Energy Physics, such as bio-medical physics and space physics, from sub-micrometer cells [8]-[10] and ray-tracing [11]-[14], up to planetary scales [15]. All Geant4-DNA developments are included in the Geant4 toolkit and benefit from the easy accessibility of the code, providing freely to the radiobiology community a wider access to such computing methods.

### 1.3 History

Historically, the Geant4-DNA project was initiated in 2001 by Dr P. Nieminen at the European Space Agency for the development of a computing platform allowing an estimation of the biological effects of ionising radiation using the Geant4 toolkit, in the perspective of future exploration missions of the solar system. A preliminary set of physics processes adapted to microdosimetry in liquid water down to the electronVolt scale was delivered into the Geant4 toolkit in December 2007 [3] and has been improved since then [16]-[18]. The project is now entirely developed and managed by the "Geant4-DNA collaboration".

### 1.4 The Geant4-DNA collaboration

The "Geant4-DNA collaboration" gathers Geant4 developers, who are members of the Geant4 collaboration, and external consultants with specific expertise, including theoretical elementary particle physics, radiolysis and microdosimetry. The current Geant4-DNA collaborators are listed as authors of this paper. The project is a full activity of the Geant4 Low Energy Electromagnetic Physics Working Group of the Geant4 collaboration and all Geant4-DNA on-going developments are included into the public releases of the Geant4 toolkit. They are fully described in a dedicated web site [19].

## 2 The Geant4-DNA physics processes and models

Physical interactions are described by specific C++ "process classes", which compute the total cross section of specific physical interactions (e.g. elastic scattering, ionisation, etc); a full description of the interaction products (kinematics, production of secondary particles, energy deposits, etc.) is provided as well. In Geant4-DNA, these processes are purely discrete, i.e. they simulate all physical interactions following a step-by-step precise tracking without using any condensation technique. The user can compute a certain physical quantity according to a variety of models (theoretical or semi-empirical) by dedicated "model classes", that may be complementary in energy ranges or fully alternative. A single "process class" can evoke one or several "model classes".

All Geant4-DNA process and model classes have been entirely redesigned for the last public version of Geant4 (9.3 BETA - June 2009), along with all classes of the Low Energy Electromagnetic Physics package of Geant4, in order to adopt a coherent approach to the modelling of all electromagnetic interactions in Geant4. The redesigned classes include new features convenient to the users, such as, for example, direct access to cross section values for a given particle energy. The new design and its features are fully described in [16]-[17].

### 2.1 Available physics processes and models

The Geant4-DNA extension set covers currently the dominant interactions of light particles and ions including electrons, protons, hydrogen, helium particles and their charged states down to the electronVolt scale in liquid water, the main component of biological matter. The corresponding models and their experimental validation are described and discussed in detail in [16]-[17]. Some of these models are purely analytical, others make use of interpolated cross section data tables for a faster computation. The list of available processes and models as they are available in the current public version of Geant4 (9.3 BETA) is given in Table 1, as well as their applicability range.

Table 1. List of Geant4-DNA physical processes available in version 9.3 BETA of the Geant4 toolkit released in June 2009. The high and low energy applicability of the corresponding models is indicated as well.

| Process | Model | Low energy limit | High energy limit |
|---|---|---|---|
| **Electrons** | | | |
| Elastic scattering | Screened Rutherford | 8.23 eV | 10 MeV |
| | Champion (alternative) | 8.23 eV | 10 MeV |
| Excitation | Emfietzoglou | 8.23 eV | 10 MeV |
| Ionisation | Born | 12.61 eV | 30 keV |
| **Proton** | | | |
| Excitation | Miller and Green | 10 eV | 500 keV |
| | Born | 500 keV | 10 MeV |
| Ionisation | Rudd | 100 eV | 500 keV |
| | Born | 500 keV | 10 MeV |
| Charge decrease | Dingfelder | 1 keV | 10 MeV |
| **Hydrogen** | | | |
| Ionisation | Rudd | 100 eV | 100 MeV |
| Charge increase | Dingfelder | 1 keV | 10 MeV |
| **He$^{2+}$** | | | |
| Excitation | Miller and Green | 1 keV | 10 MeV |
| Ionisation | Rudd | 1 keV | 10 MeV |
| Charge decrease | Dingfelder | 1 keV | 10 MeV |
| **He$^{+}$** | | | |
| Excitation | Miller and Green | 1 keV | 10 MeV |
| Ionisation | Rudd | 1 keV | 10 MeV |
| Charge increase | Dingfelder | 1 keV | 10 MeV |
| Charge decrease | Dingfelder | 1 keV | 10 MeV |
| **He (neutral helium)** | | | |
| Excitation | Miller and Green | 1 keV | 10 MeV |
| Ionisation | Rudd | 1 keV | 10 MeV |
| Charge increase | Dingfelder | 1 keV | 10 MeV |

Fig. 1 to Fig. 4 show the total cross section of all physics processes and corresponding models available in the Geant4-DNA extension of Geant4 9.3 BETA release (June 2009). Some of these models are currently being refined and are expected to be available in the next release (version 9.3) of the Geant4 toolkit in December 2009 [16]. The validation of the cross section models is the object of another paper by the Geant4-DNA collaboration [17].

It should be noted here that the excitation and charge increase/decrease processes are specific to the Geant4-DNA extension, and are not available in the Standard or in the Low Energy electromagnetic package of Geant4.

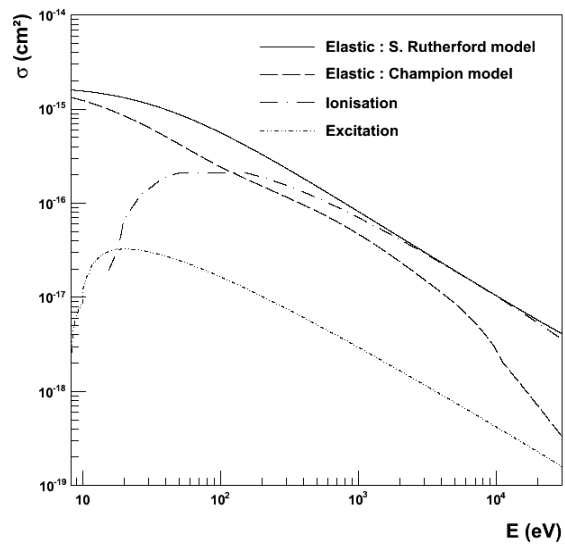

Fig. 1. Total cross section of electron processes available in Geant4-DNA 9.3 BETA as a function of electron incident energy, ranging from 8.23 eV to 30 keV: elastic scattering (Screened Rutherford model in solid line, Champion's model in dashed line), ionisation (dashed-dotted line) and excitation (dashed-dotted-dotted line). Elastic scattering is the dominant process in the lower energy range, i.e. below 100 eV. Elastic scattering (Screened Rutherford model only) and ionisation dominate at higher energies. A description of both these elastic models can be found in [16] and [18].

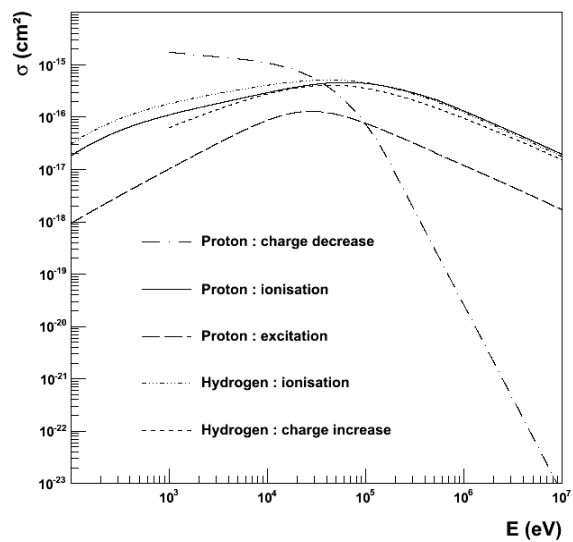

Fig. 2. Total cross section of proton and hydrogen processes available in Geant4-DNA 9.3 BETA as a function of incident particle energy (100 eV to 10 MeV). Protons: charge decrease (dashed-dotted line), ionisation (solid line), excitation (long dashed line). Hydrogen : ionisation (dashed-dotted-dotted line), charge increase (short dashed line). For protons, charge decrease dominates at low energy while ionisation becomes dominant at higher energies. For hydrogen, ionisation dominates in the whole energy range.

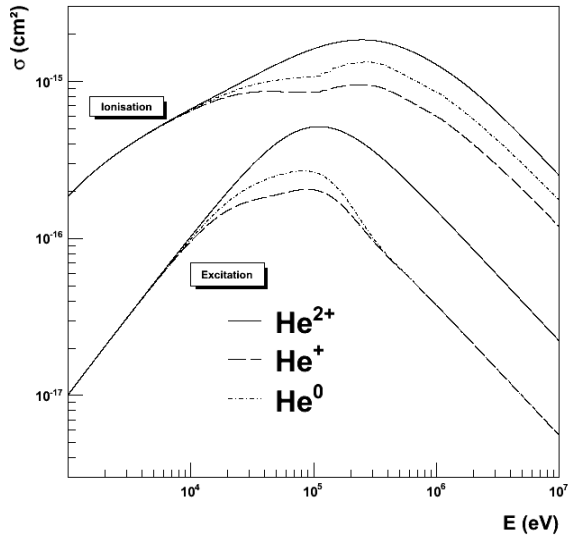

Fig. 3. Total cross section of neutral helium (dashed-dotted line), He$^+$ (dashed line) and He$^{2+}$ (full line) ionisation and excitation processes available in Geant4-DNA 9.3 BETA as a function of incident energy (from 1 keV to 10 MeV). Ionisation is the dominant process at all energies.

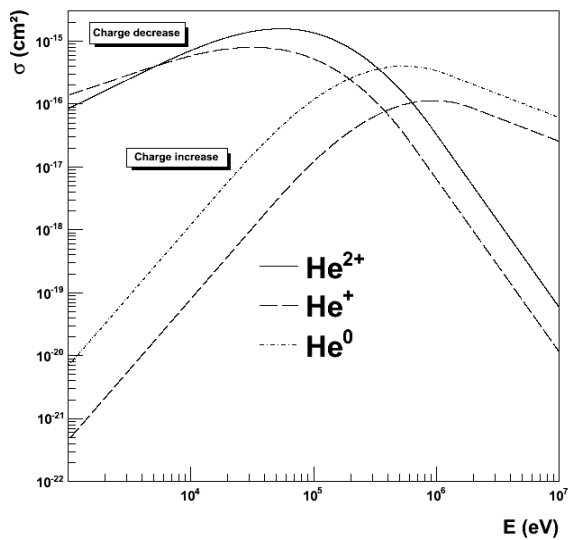

Fig. 4. Total cross section of neutral helium (dashed-dotted line), He$^+$ (dashed line) and He$^{2+}$ (solid line) charge increase and charge decrease processes available in Geant4-DNA 9.3 BETA as a function of incident energy, ranging from 1 keV to 10 MeV. At low and high energies, the ionisation process (shown in Fig. 3) remains the dominant process.

### 2.2 Implementing a Geant4-DNA "physics list"

In a user application, the Geant4 user must specify in a dedicated "physics list" class the particles and the corresponding physical processes affecting these particles (for example electrons are sensitive to the elastic scattering, excitation and ionisation processes). The default Geant4-DNA physics list, containing all particles and processes given in Table 1 and shown in Fig. 1 to Fig. 4, can be found in the "microdosimetry advanced example" (see more details in section 5), as well as in the Geant4-DNA web site [19].

### 2.3 Track structure modelling

As an illustration of the nanometer-scale tracking capabilities of the Geant4-DNA extension, we show in Fig. 5 and Fig. 6 the track structures of several particles obtained with all processes listed in Table 1 .

Fig. 5 shows the projected ionising track structure of a single 1 keV electron in liquid water. The incident primary electron and the secondary electrons produced by ionisation are tracked down to the lowest energy limit of 8.23 eV. Below this energy, the electrons are stopped and deposit their energy in the medium. The dotted cloud represents each single elastic scattering interaction. In this energy range, elastic scattering largely dominates excitation (magenta circles) and ionisation (red starts), as shown in Fig. 1.

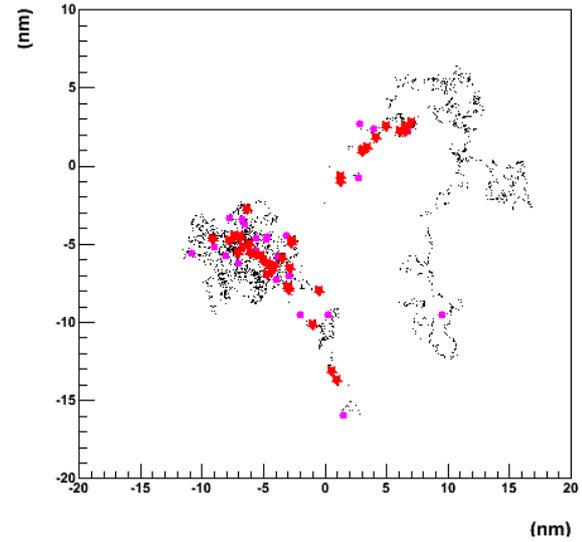

Fig. 5. Projected 2D pattern generated by a single 1 keV electron in liquid water using the Geant4-DNA physics processes. The primary particle originates at the (0,0) position. Elementary interactions are shown: elastic scattering using Champion's theoretical model [18] (black dots), excitation using Emfietzoglou's model (magenta circles) and ionisation according to Born's model (red stars). See references [16] and [18] for more details about the models.

Fig. 6 shows a comparison of 3D track structures obtained for all particles and processes available in the Geant4-DNA extension. The primary particles have an incident kinetic energy of 10 keV and are emitted towards the positive z direction (refer to the figure legend for particle identification).

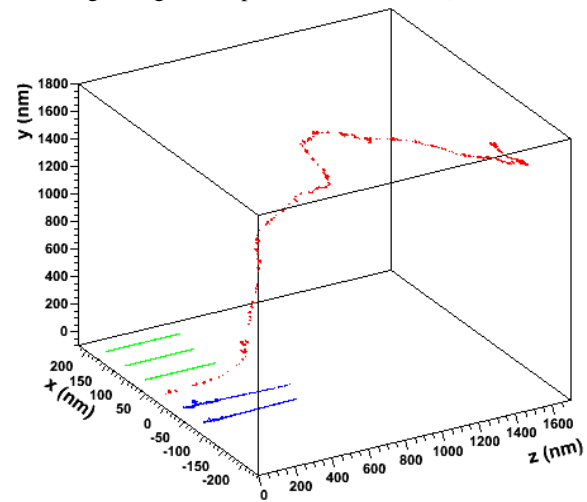

Fig. 6. Comparisons of six 3D track structures obtained with Geant4-DNA physics processes for single 10 keV incident particles in liquid water. The particles are emitted towards the positive z direction and from different x positions for the sake of clarity: proton (x=-50 nm), hydrogen (x=-100 nm), electron (x=0), He$^{2+}$ (x=50 nm), He$^+$ (x=100 nm), helium (x=150 nm).

Involved particles are tracked losing their energy step-by-step until they reach the energy limits given in Table 1. Below these limits, the particles are stopped and deposit their energy locally. Each point in the track structure corresponds to a single physical interaction.

Fig. 5 and Fig. 6 can be easily reproduced using the Geant4 "microbeam advanced example", which is fully described in section 5 of this paper.

### 2.4 Step size limitation

In Geant4 the length of the simulation steps is directly derived from the computation of the total cross section of each physics process affecting the current particle. The size of the step can however be artificially limited to a maximum value when precise particle tracking is needed in specific volumes of the geometry. For this purpose, a dedicated discrete process, called *G4StepLimiter* [20], can be added to the list of processes affecting a particle type in the user "physics list". This process allows the user to apply different maximum step size values to different volumes of the geometry. This artificial step limitation does not affect the physical interactions involved (shown for example in Fig. 5 for 1 keV electrons); in other words, the track structure of particles is independent of the user step size limit. This is shown in Fig. 7 for $10^3$ electrons of 1 keV in liquid water. The red curve shows the ratio of the sum of the number of steps computed from the physics processes and the number of additional steps forced by the *G4StepLimiter* process to the number of steps computed from the physics processes only, as a function of the user step size limit. The smaller the user step size the higher the ratio. The usage of too short user step size limits leads inevitably to an increase of the simulation computing time. As the user step size limit increases, the ratio approaches unity, since the physics processes only limit the size of the steps. The blue curve shows the relative change of the full electron physical track structure obtained with and without the G4StepLimiter process, as a function of the user step size limit. It remains constantly equal to one independently of the value of the user step size limit. The same phenomenon characterises all the other particles and corresponding Geant4-DNA processes. Therefore, no user defined step limitation is required for their use.

### 2.5 Adjusting the low energy limit of tracked particles

In the most recent public release of Geant4 (version 9.3 BETA, released in June 2009), Geant4-DNA physics processes are applicable to particles down to an energy limit of 8.23 eV for electrons, corresponding to the lowest excitation level of the liquid water molecule. Protons and hydrogen are tracked down to 100 eV, while helium particles and their charged states down to 1 keV. Such limits may however be lowered in the future developments of the toolkit. However, a Geant4 user has the possibility to raise these values if they are not interested in the tracking of a given particle type down to such low energies. The computing time needed for the simulation will decrease consequently. Below the specified energy threshold, the particle tracking will stop and the particle will deposit its energy locally in the target medium. For this purpose, similarly as for the step limitation case, the user can add a dedicated discrete process, named *G4UserSpecialCuts* [20], to the list of processes affecting the particles involved in the simulation described in the user "physics list". Again, the low energy limit value is specified when constructing the simulated geometry and can be applied to selected volumes as required. For convenience to users, this method is described in more details in the Geant4-DNA web site [19].

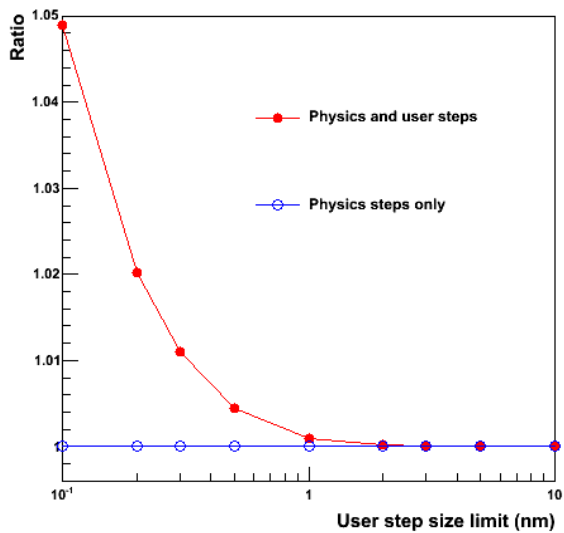

fig. 7. Dependence of the ratio of the number of physics steps and user steps to the number of physics steps only (red curve) as a function of the user step size limit expressed in nanometres. The blue curve illustrates that the physical track structures do not depend on the user step size limit. These results have been obtained for $10^3$ electrons of 1 keV in liquid water using Geant4-DNA physics processes.

## 3 Modelling direct radiation damages to DNA

As the Geant4-DNA physics processes and models are fully integrated into the Geant4 toolkit, they can be easily combined with Geant4 geometry modelling capabilities. In particular, it becomes possible to implement the geometry of biological targets with a high resolution at the sub-micrometre scale and fully track particles within these geometries using the Geant4-DNA physics processes. These geometries represent a significant improvement of the geometrical models used so far for dosimetry studies with the Geant4 toolkit at the biological cell scale [9]. Two approaches may be followed in order to implement high resolution geometries: the voxellized approach and the atomistic approach.

### 3.1 Voxellized biological geometries

The voxellized approach, based for example on reconstructed 3D images acquired with high resolution imaging techniques such as confocal microscopy, was presented in [10]. We developed high resolution cellular phantom geometrical models for Geant4 at the sub-micrometre scale. These phantoms represent realistic individual human keratinocyte cells - including the cell nucleus, inner nucleoli and cytoplasm - for a more precise dosimetry calculation in cellular irradiation experiments involving single-ion microbeams. The physics calculations were performed using the Geant4 low energy electromagnetic processes. The Geant4 "microbeam" advanced example shows how to implement such cellular phantoms, and is already available to users directly in the Geant4 toolkit. It is fully documented in the Geant4 "advanced examples" web site [21].

### 3.2 An atomistic DNA geometrical model

In the atomistic approach, we propose to model higher granularity biological targets at the nanometre scale, such as the DNA molecule, using combinations of simple mathematical volumes. The DNA geometrical model presented in this paper is inspired from the work of M. A. Bernal and J. A. Liendo [22] in their investigation of the capabilities of the PENELOPE Monte Carlo code in nanodosimetry. This model is organized into four geometry levels: the deoxynucleotide pairs, the DNA double helix, the nucleosomes (two DNA loops wrapped around a chromosomal protein named histone) and the chromatin fibre (i.e. the DNA assembled into chromosomes). All dimensions and positions correspond to the B-DNA molecular configuration and they are fully described in [22].

The implementation of this geometrical model into a Geant4 application can be easily achieved following a "top to bottom" order, from the whole chromatin fibre down to the DNA bases, benefiting from the geometrical symmetries of the model. Thanks to these symmetries, it is not necessary to implement into Geant4 each single deoxynucleotide pair, a tedious and time-consuming task which would considerably slow down the simulation and lead to an inefficient navigation in the elementary geometrical volumes. Instead, individual bases are modelled as sectors of cylindrical shells, with an inner radius of 0.5 nm and an outer radius of 1.185 nm. They have a thickness of 0.33 nm, in the sub-nanometre range. The positions of 100 pairs of bases are parameterised into a DNA helix loop. Two of these loops are wrapped around a histone to assemble a nucleosome, as shown in Fig. 8.

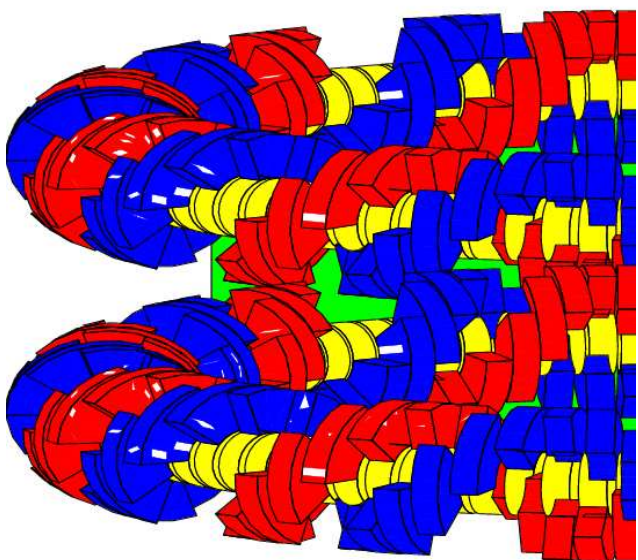

Fig. 8. Close-up of the Geant4 visualisation of a half-nucleosome described above, showing the two DNA loops. Each DNA loop contains a total of 100 base pairs. The yellow cylinders represent the base pairs and the red and blue sectors are the corresponding opposed phosphodiester groups. The distance between the two loops has been increased for better visibility. The nucleosome has a height of 5.2 nm and a diameter of 10.5 nm. This figure was obtained using the DAWNFILE visualization driver available in the Geant4 toolkit. In this figure some volumes seem to overlap, however this is simply an artefact of the 3D visualization.

Each "slice" of the chromatin fibre is made of 6 nucleosomes, and are wrapped in a compact configuration, that is replicated along the fibre length. For example, Fig. 9 shows a chromatin fibre made of 10 "slices". Such a fibre has a 30 nm diameter, a length of about 100 nm and it contains $1.2 \times 10^4$ DNA base pairs. All these geometrical volumes are filled with liquid water since Geant4-DNA processes are defined for liquid water only; we can anyhow consider such an approximation quite realistic. Note that both figures were obtained using the Geant4 DAWNFILE visualization driver available in the toolkit [20] and able to display sub-nanometre geometrical volumes.

Particles can be ejected onto the fibre and a user step limit of 0.1 nm, smaller than the smallest volume size contained in the geometrical model, can be chosen in order to reach a detailed tracking in the smallest volumes of the geometry (0.33 nm in our case). Each elementary energy deposit occurring in the particle shower can be precisely located and recorded within the geometry. Bases are easily identified according to the slice, nucleosome and DNA helix loop they belong to. Consequently, it is possible to estimate how many DNA single strand breaks (SSB) and double strand breaks (DSB) can be induced within the chromatin fibre by the incident ionising particle. Such strand breaks are one of the simplest biological damage observables.

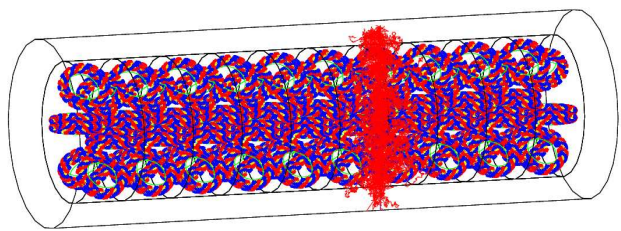

Fig. 9. Geant4 visualisation of the whole chromatin fibre described in the text and irradiated by a single 500 keV He$^+$ particle, emitted perpendicularly to the main revolution axis of the fibre. DNA phosphodiester groups are represented in blue and red. They are assembled into helix loops. Each histone has two DNA loops and forms a nucleosome. Each of the 10 "slices" of this fibre contains 6 nucleosomes (shown in Fig. 8) in the B-DNA conformation. The full track structure induced by the incident particle and simulated with Geant4-DNA physics processes is also shown. This figure was obtained using the DAWNFILE visualization driver available in the Geant4 toolkit.

The shower of particles shown in Fig. 9 has been obtained by emitting a single 500 keV He$^+$ particle onto the fibre. A SSB is counted if one target (phosphodiester group) receives an energy deposit greater than 10.79 eV [22], the first ionisation level of the liquid water molecule. A DSB is counted if two SSBs are located in opposite strands, and the breaks are distant by less or equal than 10 base pairs. As an illustration, with this definition, the Geant4 simulation predicts that a single 500 keV He+ particle could for example generate 9 SSBs and 2 DSBs in the fibre geometrical model, resulting from the direct interaction between ionising particles and the geometrical model.

Undoubtedly, the presented geometrical model may be refined by increasing the granularity of the model (e.g. including an atomic description of phosphodiester groups) and simulations involving higher statistics should be performed before drawing any quantitative conclusion regarding the validity of such DNA single and double strand breaks estimations with the Geant4-DNA extension. Nevertheless, this study demonstrates the promising potentialities offered by the usage in tandem of Geant4-DNA physics processes with high resolution geometrical models at the nanometre scale.

## 4 Modelling chemistry with Geant4-DNA

Ionising radiation can directly damage the DNA molecule if energy loss events, such as ionisation, transfer a sufficient energy amount to the molecule able to generate strand breaks. This stage usually takes place between the instant when the primary ionising particle penetrates into the biological cell and lasts about $10^{-15}$ s [23]. However, it is well known that DNA damages resulting from the direct interaction of ionising radiation with the DNA molecule do not represent the dominant mechanism for DNA damage when low-LET radiation is involved. Most of the damages are indeed induced to the DNA molecule by indirect effects of ionising radiation, occurring between $10^{-15}$s up to $10^{-12}$s. During this stage, water molecules that have been excited and ionised before $10^{-15}$s may de-excite and dissociate in order to generate molecular radical species. These radicals are reactive species, they diffuse in the DNA surrounding medium interacting with water molecules and DNA, and they are responsible for most of the damages caused to the DNA molecule.

The implementation of reactive radical species in the Geant4 toolkit is currently in progress. Geant4 is able to model particles interaction with matter but cannot model interactions between particles, such as collective effects. We expect to deliver specific classes able to handle the modelling of radical species and their interactions in the near future, bringing additional functionality to the Geant4 toolkit.

## 5   The microdosimetry example for Geant4 users

The Geant4 toolkit includes an "advanced example" named "microdosimetry advanced example". The list of physics processes ("physics list") available in this example has been adapted in Geant4 version 9.3 BETA (release of June 2009) to the new design of Geant4-DNA processes and models, which is now common to Geant4 Standard and Low Energy electromagnetic processes and models. It demonstrates to users how to implement the Geant4-DNA physics list (see Table 1) including the possibility to identify Geant4-DNA physics processes by user-defined names, generate primary particles and extract the full 3D track structures of particles in liquid water using the entire set of Geant4-DNA processes. A ROOT [24] macro file is also provided in order to easily display the track structure of tracked particles, such as the tracks shown in Fig. 5 and Fig. 6. This example is fully described in the Geant4 Advanced Examples working group web site [21].

## 6   Geant4 for VMware™

All Geant4-DNA physics processes and models are entirely available in the low energy electromagnetic package of the Geant4 toolkit and are fully accessible to Geant4 users. In order to facilitate the installation of Geant4 to novice users, we propose free of charge and licensing a full software suite ready-to–use for Windows™ or Macintosh™ users running under the VMware™ virtualization environment. After download, the user will emulate a complete Scientific Linux™ machine where the full and most recent version of Geant4 (including the Geant4-DNA extension and the cited examples) publicly available from the Geant4 web site [7] and associated tools are ready-to-use, without any additional software installation needed. In particular, all necessary software allowing data collection (e.g. histograming) and analysis (e.g. ROOT) is already configured to be used in any Geant4 user application. This service is proposed by C. Seznec and I. Moreau at the Centre d'Etudes Nucléaires de Bordeaux-Gradignan, France, and is available through the Internet [25]. It has already been conveniently used for several Geant4 tutorials [26]. User support is kindly provided if needed.

## 7   Conclusion and perspectives

The current status and developments available in the Geant4-DNA extension of the Geant4 toolkit have been presented in this paper. Geant4-DNA processes extend the Geant4 toolkit to a sub-micrometer modelling of particle physical interactions in liquid water. In combination with preliminary detailed geometrical models of the DNA molecule, we expect to be able to simulate direct damages such as SSBs and DSBs induced by ionising radiation.

Future developments will include the refinement and the inclusion of additional physics models for photons, for heavier ions like carbon and oxygen (a common requirement of novel hadron therapy techniques and of space radiation applications) and for other target materials of biological interest (e.g. DNA bases). These new models will cover extended energy ranges in combination with already existing Geant4 physics models (standard electromagnetic processes and low energy electromagnetic processes). In particular, the inclusion of vibrational excitation models from the work of Michaud and Sanche down to 0.025 eV is planned.

The modelling of chemistry processes, in particular for the production of oxidative radical species, their diffusion and their mutual interaction is planned and would allow for the simulation of non-direct DNA damages.

The development of high resolution geometrical models at the molecular scale, using both the voxellized approach and/or the atomistic one is still an open requirement.

Geant4-DNA predictions will be compared and validated with experimental data available on dedicated facilities, such as single-ion microbeams for single-cell irradiation allowing a precise quantification of DNA damages (SSBs, DSBs) after irradiation. The study of these damages is important, as they represent the earliest ones, that occur before any complex cellular response and repair mechanisms are activated.

Additionally, comparisons to other simulation codes able to model track structures of particles and chemical species at the sub-micrometer level as well as predict DNA damages are foreseen.

The project will benefit from the open-source accessibility of the Geant4 toolkit and from its international visibility, since it is a full component of Geant4. It will provide freely and in complete transparency a fully integrated simulation platform of DNA damages of radiation to any interested user. The project is expected to have an impact in several research domains, beyond the initial scope of Geant4 for High Energy Physics (HEP) simulations. Indeed, the validated models would be able to cover application domains involving ionising radiation and biological entities, at the cell scales. Foreseen applications include, for example, radioprotection and microdosimetry for long duration space exploration programs involving humans and biological samples, as well as radiotherapy applications.

The Geant4-DNA project will be partly funded by the French Agence Nationale de la Recherche (ANR) for the 2010-2013 period. The ANR driving and funded partners are the Centre d'Etudes Nucléaires de Bordeaux-Gradignan / University Bordeaux 1 / Centre National de la Recherche Scientifique (CNRS) / Institut National de Physique Nucléaire et de Physique des Particules (IN2P3) in Gradignan (France), the Laboratoire de Radiolyse / Commissariat à l'Energie Atomique (CEA) in Saclay (France), and the Laboratoire de Physique des Collisions Moléculaires / University of Metz (France). The Geant4-DNA collaboration is open to any interested collaborator.

### Acknowledgments

The authors wish to thank the organizers of the Asia Simulation Conference 2009 for their kind invitation to present this work about the Geant4-DNA project and collaboration.

We also express our deep gratitude to the Geant4-collaboration members for their constant support and guidance, and especially to Dr Makoto Asai (SLAC National Accelerator Laboratory, Menlo Park, CA, USA), to Dr Vladimir Ivantchenko and to Dr Gabriele Cosmo (European Organization for Nuclear Research, Geneva, Switzerland).

The French Région Aquitaine funded the computing farm based at CENBG dedicated to the Geant4-DNA project that will start to operate from autumn 2009.

Our attendance to the Asia Simulation Conference 2009 was partly supported by the France-Japan Particle Physics Laboratory (FJPPL) between CNRS/IN2P3 and CEA in France and the High Energy Accelerator Research Organization (KEK) in Japan. In particular, this laboratory has been promoting Geant4 developments at the physics-medicine-biology frontier since 2007 [27].

### References

[1] A. M. Kellerer, Invisible threat, IAI2001, NUPECC, 2001
[2] K. H. Chadwick, H. P. Leenhouts, A molecular theory of cell survival, Phys. Med. Biol., vol. 18, pp. 78-87, 1973
[3] S. Chauvie, Z. Francis, S. Guatelli, S. Incerti, B. Mascialino, P. Moretto, P. Nieminen, M. G. Pia, Geant4 physics processes for microdosimetry simulation: design foundation and implementation of the first set of models, IEEE Trans. Nucl. Sci. 54 ,6-2, 2619-2628, 2007

[4] W. Friedland, P. Jacob, H.G. Paretzke, M. Merzagora, A. Ottolonghi, Simulation of DNA fragment distributions after irradiation with photons, Rad. Env. Bio. 38, 39-47, 1999
[5] S. Agostinelli *et al.*, Geant4 – a simulation toolkit, Nucl. Instr. and Meth. A 506, 250-303, 2003.
[6] J. Allison *et al.*, Geant4 developments and applications, IEEE Trans. Nucl. Sci. 53, 1, 270-278, 2006
[7] Geant4 collaboration web site: http://cern.ch/geant4
[8] S. Incerti, Ph. Barberet, R. Villeneuve, P. Aguer, E. Gontier, C. Michelet-Habchi, Ph. Moretto, D.T. Nguyen, T. Pouthier and R.W. Smith, Simulation of cellular irradiation with the CENBG microbeam line using Geant4, IEEE Trans. Nucl. Sci., 51, 4, 1395-1401, 2004
[9] S. Incerti, N. Gault, C. Habchi, J.L. Lefaix, Ph. Moretto, J.L. Poncy, T. Pouthier and H. Seznec, A comparison of cellular irradiation techniques with alpha particles using the Geant4 Monte Carlo simulation toolkit, Rad. Prot. Dos. 122, 1-4, 327-329, 2006.
[10] S. Incerti, H. Seznec, M. Simon, Ph. Barberet, C. Habchi, Ph. Moretto, Monte Carlo dosimetry for targeted irradiation of individual cells using a microbeam facility*,* Rad. Prot. Dos. 133, 1, 2-11, 2009.
[11] S. Incerti, R.W. Smith, M. Merchant, G.W. Grime, F. Méot, L. Serani, Ph. Moretto, C. Touzeau, Ph. Barberet, C. Habchi and D.T. Nguyen, A comparison of ray-tracing software for the design of quadrupole microbeam systems, Nucl. Instrum. and Meth. B 231, 76-85, 2005
[12] S. Incerti, C. Habchi, Ph. Moretto, J. Olivier and H. Seznec, Geant4 simulation of the new CENBG micro and nanoprobes facility, Nucl. Instrum. and Meth. B 249, 738-742, 2006
[13] S. Incerti, Q. Zhang, F. Andersson, Ph. Moretto, G. W. Grime, M. J. Merchant, D. T. Nguyen, C. Habchi, T. Pouthier, H. Seznec, Monte Carlo simulation of the CENBG microbeam and nanobeam lines with the Geant4 toolkit, Nucl. Instrum. and Meth. B 260, 20-27, 2007
[14] F. Andersson , Ph. Barberet , S. Incerti , Ph. Moretto , H. Seznec , M. Simon, A detailed ray-tracing simulation of the high resolution microbeam at the AIFIRA facility, Nucl. Instrum. and Meth. B 266, 1653-1658, 2008
[15] A. Le Postollec, S. Incerti, M. Dobrijevic, L. Desorgher, G. Santin, Ph. Moretto, O. Vandenabeele-Trambouze, G. Coussot, L. Dartnell, P. Nieminen,Monte-Carlo Simulation of the radiation environment encountered by a biochip during a mission to Mars, Astrobiology 9 (3) 311-323, 2009
[16] Z. Francis, S. Incerti, R. Capra, B. Mascialino, C. Villagrasa, G. Montarou, A set of processes dedicated for the molecular scale track structure simulations in liquid water for the Geant4 Monte-Carlo toolkit, submitted to Applied Radiation and Isotopes (2009).
[17] S. Incerti, B. Mascialino, C. Champion, H. N. Tran, R. Capra, Z. Francis, S. Guatelli, P. Guèye, A. Mantero, P. Moretto, P. Nieminen, A. Rosenfeld, C. Villagrasa, C. Zacharatou, Statistical validation of Geant4-DNA physics models for microdosimetry, in preparation, 2009
[18] C. Champion, S. Incerti, H. Aouchiche, D. Oubaziz, A free-parameter theoretical model for describing the electron elastic scattering in water in the Geant4 toolkit, Rad. Phys. Chem. 78, 745-750, 2009
[19] The Geant4-DNA web site is available from the Geant4 Low Energy Electromagnetic working group web site, which is accessible directly from the Geant4 collaboration web site [7].
[20] Geant4 User's Guide For Application Developers, available from [7]
[21] The Geant4 "advanced examples" web site: http://geant4advancedexampleswg.wikispaces.com/
[22] M. A. Bernal and J. A. Liendo, An investigation on the capabilities of the PENELOPE MC code in nanodosimetry, Med. Phys. 36, 620-625, 2009.
[23] W. Friedland, P. Jacob, H. G. Paretzke, M. Merzagora, A. Ottolenghi, Simulation of DNA fragment distributions after irradiation with photons, Radiat Environ Biophys 38, 39–47, 1999
[24] The ROOT web site: http://root.cern.ch
[25] The Geant4 for VMware™ software is available in free download directly from the following web site: http://geant4.in2p3.fr/rubrique.php3?id_rubrique=8&lang=en or from [26]
[26] Geant4@IN2P3 web site: http://geant4.in2p3.fr
[27] France-Japan Particle Physics Laboratory (FJPPL) web site: http://fjppl.in2p3.fr